\documentclass{iau}

\usepackage{amsmath}
\usepackage{graphicx}
\usepackage{multirow}
\usepackage{csquotes}

\newcommand\Hzsns{Hz~s$^{-1}$}

\begin{document}

\lefttitle{Margot}
\righttitle{Advancing the Search for Technosignatures}

\jnlPage{1}{7}
\jnlDoiYr{2021}
\doival{10.1017/xxxxx}

\aopheadtitle{Proceedings IAU Symposium}
\editors{J. Haqq-Misra \& R. Kopparapu, eds.}

\title{Results of ten years of UCLA SETI searches with the Green Bank Telescope}

\author{Jean-Luc Margot}
\affiliation{University of California, Los Angeles, Los Angeles, CA, USA}

\begin{abstract}
  We have been conducting a search for narrowband radio signals with the L-band receiver (1.15-1.73 GHz) of the 100 m diameter Green Bank Telescope \citep{marg23setiothers}.  So far, we have captured radio emissions from 70,000+ stars and planetary systems in the $\sim$9 arcminute beam of the telescope. Our data-processing pipeline has a demonstrated 94\%-99\% efficiency for the detection of narrowband signals across the full range of frequency drift rates ($\pm$9 Hz/s).  All 100 million candidate signals detected to date were either automatically (99.5\%) or visually (0.5\%) confirmed to be anthropogenic in nature. These results allow us to place stringent limits on transmitter prevalence: at the 95\% confidence level, the fraction of stars within 20,000 ly that host a transmitter that is detectable in our search (EIRP $> 5\times 10^{16}$~W) is $<$6.3 $\times 10^{-5}$. Our most interesting signals have been uploaded to a citizen science platform (http://arewealone.earth), where 40,000+ volunteers to date have contributed insights and classifications.  We are using artificial intelligence (AI) to accelerate our search, automatically excise radio frequency interference, and improve signal detection.  UCLA SETI research has involved $\sim$200 undergraduate and $\sim$20 graduate students so far.
\end{abstract}


\maketitle

\section{Introduction}
Analysis of Kepler Mission data suggests that the Galaxy includes
billions of Earth-size planets in the habitable zone of their host
star~\citep[e.g.,][]{brys21others}.  The possibility that intelligent and
communicative life forms developed on one or more of these worlds
behooves us to conduct a search for extraterrestrial intelligence, as
proposed by \citet{Cocconi1959}, initiated by \citet{Drake1961}, and
studied in detail by \citet{cyclops}.

Technosignature searches are logical extensions of the search for
biosignatures that offer important cost and search volume advantages.
Biosignature missions cost billions of dollars (e.g., $\sim$\$10B each for
Mars Sample Return and Habitable World Observatory) and
robust technosignature searches can be conducted far more economically \citep[e.g.,][]{marg19astro2020}.
All targets presently envisioned in
biosignature
searches are located in the solar neighborhood, whereas radio
technosignatures are detectable throughout the entire
Galaxy~\citep[e.g.,][]{marg18setiothers}.  Roughly speaking, the
accessible search volume for radio technosignatures is a million times
larger than the search volume of all biosignature searches
contemplated at this time, $V_{\rm techno} / V_{\rm
  bio} > 10^6$.  In terms of the number of accessible search targets,
the ratio is approximately $N_{\rm techno} / N_{\rm bio} > 10^9$.

\section{Narrowband Radio Signals}
Narrowband radio signals
provide compelling evidence for technology because astrophysical
settings cannot generate signals with spectral purity smaller than
$\delta \nu/\nu \sim 10^{-6}$, where $\nu$ is the frequency and
$\delta \nu$ is the signal bandwidth~\citep{Tarter2001}.  For this
reason, UCLA SETI conducts L-band searches (1.15--1.73~GHz) with 3~Hz
frequency resolution, approaching $\delta \nu/\nu \sim 10^{-9}$.  If
one defines narrowband at a level that rules out a natural origin, the
1977 WOW!  signal \citep[e.g.,][]{Mendez2025others}, with $\delta
\nu/\nu \sim 10^{-5}$, is not considered narrowband.

UCLA SETI detections are not materially affected by
scintillation-induced spectral broadening over 0--25 kpc distances
\citep{Cordes1991}.  Likewise, the current focus on solar-type
(G-type) stars makes UCLA SETI detections mostly immune to spectral
broadening by the interplanetary medium \citep{gajj26}.  However,
intensity scintillations do affect detectability for lines of sight
outside the plane of the Galaxy.  At galactic latitudes less than
$\sim$1 degree, the scintillation times are short enough that
scintillations are quenched and signals remain detectable at the
nominal detection threshold.  At larger galactic latitudes, the
amplitude modulation can reach 100\%, and detection at the nominal
threshold has a probability of $\sim$~30\%
\citep[][Fig. 5]{Cordes1991}.  To ensure a 90\% probability of
detection, the signal must exceed the nominal detection threshold by a
factor of $\sim$10.

\section{Radio Beacons}
Powerful monochromatic emissions are routinely used in scientific and planetary
defense investigations with terrestrial radar systems.  For example,
I have emitted 85+ hours of 450~kW monochromatic signals with
a 70~m antenna at 8.56~GHz for scientific investigations over the past
$\sim$20 years.  Likewise, the detection of 1000+ asteroids with radar
since 1990 required monochromatic transmissions at 2.38~GHz or
8.56~GHz in virtually every instance.

It is of course not a given that other civilizations use monochromatic
radar systems.  Even if they do, the probability of detecting
extraterrestrial, accidental radar or radio transmissions is
small. Indeed, SETI is most likely to be successful if beacons exist
and if these beacons are easily recognizable.
The existence of such beacons cannot be assumed nor
rejected.
Although narrowband emissions are poor encoders of information, they
are an excellent choice for beacons, as recognized in the Cyclops
report~\citep{cyclops}.  If an extraterrestrial narrowband beacon is
detected, we will almost certainly witness the launch of substantial
efforts to search for nearby broadband
signals~\citep[e.g.,][]{mess13,lesy19}.

\section{Search Metrics}
\label{sec-metrics}
UCLA SETI has conducted searches in the Kepler and TESS fields and along the galactic plane, whereas recent efforts have focused on
targeting solar-type (G-type) stars (Table~\ref{tab-stats}).  After 2021, UCLA SETI transitioned from reporting the number of primary stars observed
to reporting the total number of known stars located within the 8.9 arcmin beamwidth of the telescope, which is the more relevant metric because the search is sensitive to any emission site within the beam~\citep{wlod20}.

\begin{table}[h]
  \begin{center}
\begin{tabular}{llrrrrr}
  Data Set & Fields & Targets & Stars & Signals & HRD  & MDFM\\
        & & (prim.) & (in beam) & ($\times10^6$) & (\# kHz$^{-1}$ hr$^{-1}$) & (Hz$^2$ m$^3$ W$^{-3/2}$) \\
  \hline
  2016     & Kepler        & 14       & 11,961 & 5.22 & 10.2 & 3.95 $\times 10^{32}$\\
  2017     & Kepler+       & 12       &  7,093 & 8.52 & 16.2 & 3.72 $\times 10^{32}$\\
  2018–19  & Gal. plane    & 30       & 26,209 & 27.0 & 24.6 & 8.47 $\times 10^{32}$\\
  2020–23  & TESS          & 62       & 12,414 & 41.2 & 18.2 & 1.75 $\times 10^{33}$\\ %
  2024-25  & G stars       & 33       & 13,180 & 26.2 & 26.5 & 9.34 $\times 10^{32}$\\
\hline                                                                            
  Total    &               & 151      & 70,857 & 105.7 &20.4 & 4.31 $\times 10^{33}$\\ %
\end{tabular}
  \end{center}
  \caption{UCLA SETI search characteristics, showing observation fields, number of primary targets, number of stars observed in the beam of the telescope, number of narrowband signals detected with S/N$>$10, hit rate density (HRD, number of detections per unit bandwidth per unit on-source time), Modified Drake Figure of Merit (MDFM).%
  }
\label{tab-stats}
\end{table}

\begin{figure}[ht!] %
  \begin{center}
    \includegraphics[width=0.8\textwidth]{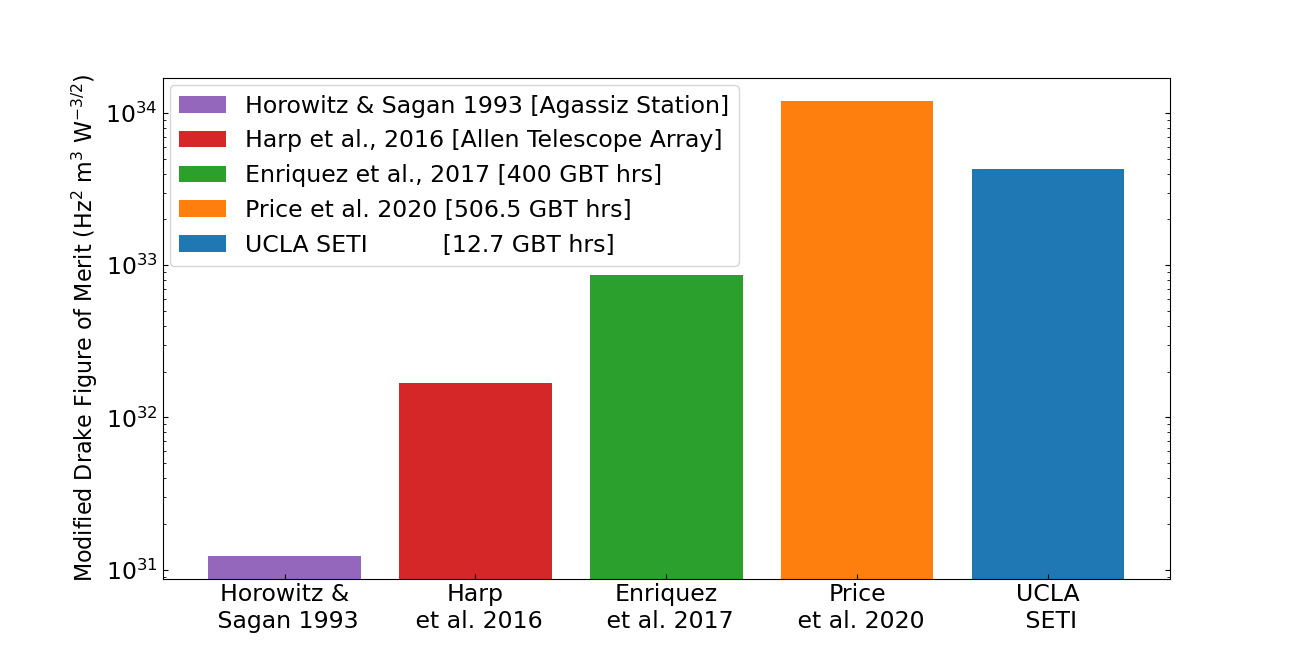} %
    \caption{ Search volume characteristics of select surveys (L-band
      component only), accounting for bandwidth, sky coverage, search
      sensitivity, pipeline efficiency, and frequency drift rate
      coverage.}
  \label{fig-dfm}
  \end{center}
\end{figure}

A key
performance metric is the {\em Modified Drake Figure of Merit} (MDFM),
which captures not only bandwidth, sky coverage, and search
sensitivity, as proposed by \citet{drak84}, but also pipeline
efficiency -- the probability of detecting a detectable signal -- and
drift range coverage -- the range of line-of-sight accelerations
sampled in a search \citep{marg23setiothers}.  In its original form,
the Drake Figure of Merit (DFM) assumes end-to-end pipeline
efficiencies of 100\%, whereas the efficiency of different programs
can vary by more than an order of magnitude
(Section~\ref{sec-inject}).  The original DFM also omits the
frequency drift rate coverage, which is an essential indicator of the
thoroughness of a search.  For instance, UCLA SETI examines drift
rates in the range $\pm8.88$ \Hzsns, corresponding to fractional drift
rates of $\pm$6.24 nHz at 1.42 GHz and maximum line-of-sight
accelerations of 1.87~ms$^{-2}$, capturing the spin and orbital
accelerations of most known exoplanets.  Because of this large drift rate coverage and
a $>$94\%
pipeline efficiency,
UCLA SETI searches outperform other searches conducted with the same
telescope, receiver, and detection threshold
\citep[e.g.,][]{enri17others, pric20others}.
Its MDFM compares favorably
to historical and modern L-band searches with larger telescope time allocations (Figure~\ref{fig-dfm}).
\nocite{Horowitz1993}
\nocite{Harp2016others}

\vspace{-1cm}\mbox{}
\section{Pipeline Efficiency}
\label{sec-inject}
An important, under-appreciated reality of SETI searches is that the
performance of detection algorithms dramatically affects estimates of
search volume (Section~\ref{sec-metrics}) and transmitter prevalence
(Section~\ref{sec-prevalence}). For this reason, we have conducted a
signal injection and recovery analysis of 10,000 synthetic signals to
quantify the end-to-end efficiency of our pipeline, i.e., the
probability that the pipeline actually detects a detectable signal.
We found that that the UCLA SETI pipeline efficiency for narrowband chirp signals is
94\%.  When regions of dense RFI are excluded, the efficiency
increases to 98.7\%.  Importantly, the distributions of recovered signal-to-noise ratios (S/Ns)
and drift rates match those of the injected population.

We repeated the analysis to simulate a widely used SETI pipeline, where
spectra are incoherently averaged by a factor of 51.  Under these
conditions, the efficiency drops to 5.7\%, 12.7\%, and 25.3\% for
signals with drift rates of $\pm8.88$ \Hzsns, $\pm4$ \Hzsns, and $\pm2$ \Hzsns, respectively, 
with S/N and drift rate distributions that do not match the injected population.

\section{Transmitter Prevalence}
\label{sec-prevalence}
We use the formalism developed by \citet{marg23setiothers} to place 
rigorous upper bounds on the fraction of star systems that host
transmitters detectable in our search.  The probability of detecting a
transmitter in each observation of a star is 
\begin{equation}
p = f_{\rm TX} \eta_P D,
\end{equation}
where $f_{\rm TX}$ is the fraction of stars in the observed sample
that host a detectable transmitter, $\eta_P$ is the end-to-end
pipeline efficiency, and $D$ is the duty cycle of the transmitter,
i.e., the fraction of time that the transmitter is beaming in Earth's
direction.

We determine an upper limit $f_{\rm TX}^{u}$ that is consistent with
obtaining zero successes in $n$ independent binomial trials at a
confidence level CL.  This value is
\begin{equation}
  f_{\rm TX}^{u} = \frac{1 - (1 - {\rm CL})^{1/n}}{\eta_P D}.
\label{eq-solve}
\end{equation}
With some simplifying assumptions (CL = 95\%, $\eta_P \simeq 1$, $D \simeq 1$,
and $n > 20$), this result reduces to the ``rule of
three'', $f_{\rm TX}^{u} \simeq 3/n$, used by
\citet[][p.\ 53]{cyclops}.

We performed cone searches with the Gaia catalog \citep{gaiadr3} to estimate the number of stars
sampled by our observations.
Based on these values and $\eta_p$ = 94\%, we can state at the 95\%
confidence level that the fraction of stars with 100\%-duty-cycle
transmitters detectable in our search is smaller than
6.3 $\times 10^{-5}$ for stars located within 20,000 ly of the Sun and 
6.6 $\times 10^{-4}$ for stars located within 1 kpc of the Sun.

\section{Sensitivity}
The high sensitivity of the GBT is conveyed by its system equivalent flux density (SEFD) of 10 Jy.
UCLA SETI scans have a duration $\tau$ = 150 s.  We analyze the data
by producing dynamic power spectra, also known as spectrograms, with a
frequency resolution $\Delta f$ $\simeq$ 3~Hz.  In the process, we sum
the powers from two polarizations.  Because we sample the voltages
with 2-bit quantization, our quantization efficiency is $\eta_Q$ =
0.8825.
Our usual detection threshold is set at a signal-to-noise ratio (S/N)
of 10.  Using these parameters in the radiometer equation, we find
that signals with flux $S_{\rm det} = 11.3 \times 10^{-26}$ W/m$^2$
are detectable.
Therefore, an Arecibo Planetary Radar (Effective Isotropic Radiated Power=$2.2\times 10^{13}$ W) located within 415~ly (127~pc)
is detectable
in our search.
Transmitters located at the galactic center are detectable provided
they emit the equivalent of 4130 Arecibos,
well within the reach of an advanced technological society.

\section{Radio Frequency Interference}
Most narrowband signals detected by our pipeline are due to radio
frequency interference (RFI).  The UCLA SETI pipeline automatically
recognizes $\sim$99.5\% of the RFI with Doppler and
directional filters.  In the former, any signal with zero Doppler
drift rate is eliminated because of the absence of line-of-sight
acceleration between the transmitter and the receiver. In the latter,
any signal that is detected in more than one direction on the sky is
eliminated because it cannot be due to a distant point source.

\section{AI Acceleration}
\citet{Pinchuk19others} documented instances where the directional
filter fails to automatically excise RFI.  Proper disposition of the
affected signals typically requires visual confirmation.  One failure
mode occurs when the joint origin of two signals observed in two separate scans
is not correctly recognized because the agreement between their
frequency or frequency drift rate estimates exceeds the tolerance of
the filter.  We implemented a convolutional neural network (CNN) to
overcome this difficulty and recognize signal associations that were
otherwise missed by the pipeline. This CNN reduces the visual
confirmation burden by an order of magnitude~\citep{Pinchuk2022}.

Certain emitters, such as global navigation satellite systems, are
responsible for the majority of RFI in our data. Because their
waveforms exhibit distinct morphologies in dynamic spectra, it is
possible for a CNN to automatically recognize these emitters with high
confidence.  \citet{li26cnnothers} implemented an RFI excision tool
based on a ResNet-like network trained on $\sim$76,000 signals,
achieving $F_1$ scores $\simeq 0.99$.  The labeling was accomplished
by tens of thousands of volunteers on the Zooniverse platform via a
participatory science collaboration called ``Are we alone in the
universe?''.

\citet{ma23others}
pioneered
the use of a deep-learning approach to
identify technosignature candidates with a $\beta$-variational
autoencoder followed by a random forest classifier.  Their model,
however, was trained exclusively on narrowband signals with constant
drift rates, and their dynamic spectra were
averaged by factors of
8 and 51 in frequency and time, respectively.
These incoherent averaging operations ruined the sensitivity of the
search for all but the smallest drift rates originally considered, as
both evaluation of the dechirping
efficiency~\citep{Margot2021setiothers} and recovery of injected
signals~\citep{marg23setiothers} demonstrate.  We have addressed these
limitations by training a similar network on synthetic signals
with a range of morphologies and by applying it to data without
averaging.
This tool allows us to detect signal types that
were not previously detectable
by our pipeline~\citep{Koron2026}.

\section{Data Archival}
To enable future archival searches, we publish the signal characteristics of all our candidate detections to an online repository~\citep[e.g.,][and subsequent datasets]{seti16datasetdoi}, including time of observation, source name, frequency, frequency drift rate, estimated bandwidth, signal-to-noise ratio, and other diagnostic flags.  The size of the repository grows by approximately 5 million entries per hour of telescope time.

\section{SETI Course}

Since 2016, I have been teaching an annual, project-based SETI course,
in which students design an observing sequence, participate in remote
observing at the GBT, analyze the data, and write code to add features
to the data-processing pipeline.  Approximately 200 undergraduate and
20 graduate students have taken the course so far.  The SETI course
helps develop skills in astronomy, signal processing,
telecommunications, computer science, and agentic coding.

\section{Funding}

By polling SETI scientists and consulting NASA award summaries, it is
possible to quantify NASA's investment in the search for
technosignatures over the years.  Between 1994 and 2024, NASA awarded
11 technosignature grants~\citep{psu}, including one to UCLA SETI
in 2021, with a total investment of \$4,189,577 in unadjusted
dollars or \$5,573,881 in inflation-adjusted dollars.

Over the same 1994--2024 period, NASA appropriations totaled \$803,925 millions of inflation-adjusted dollars~\citep{tps}.
Therefore, NASA allocated 0.0007\% of its 1994-2024 appropriations to the search for technosignatures.

In 2022, NASA changed the scope of its Exoplanet Research Program (XRP) solicitation.  Whereas the 2020 solicitation had stipulated that ``Observational, archival, theoretical, and modeling proposals focused upon the detection of technosignatures are in scope of the XRP'', the 2022 solicitation added the following restriction: ``[but] only if the proposal convincingly demonstrates that the focus of the investigation is the advancement of exoplanet science''.  Because SETI is not inherently focused on advancing exoplanet science, the newly added restriction severely limit opportunities for federal funding.

UCLA SETI functions primarily through the generous support of The Queen's Road Foundation, 
Robert Meadow and Carrie Menkel-Meadow, Larry Lesyna, Joe and Andrea
Straus, the Rattray Kimura Foundation, Michael Thacher and Rhonda
Rundle, Janet Marott, Arnie Boyarsky, and other generous donors. We
are extremely grateful for their support.

\pagebreak
\bibliographystyle{iaulike} 
\bibliography{bib.bib}

\end{document}